\pdfoutput=1 
\documentclass{JINST}
\graphicspath{{./figures/}}

\title{The Brighter-Fatter and other Sensor Effects in CCD Simulations
for Precision Astronomy}

\author{C.W. Walter\\
Department of Physics,  Duke University \\
Durham, NC 27708, USA \\
E-mail: \email{chris.walter@duke.edu}}

\abstract{ Upcoming and current large astronomical survey experiments
  often seek to constrain cosmological parameters via measurements of
  subtle effects such as weak lensing, which can only be measured
  statistically.  In these cases, instrumental effects in the image
  plane CCDs need to be accounted and/or corrected for in measurement
  algorithms.  Otherwise, the systematic errors induced in the
  measurements might overwhelm the size of the desired effects.
  Lateral electric fields in the bulk of the CCDs caused by field
  shaping potentials or space charge build up as the electrons in the
  image are acquired can cause lateral deflections of the electrons
  drifting in the CCD bulk.  Here, I report on the LSST effort to
  model these effects on a photon-by-photon basis by the use of a
  Monte Carlo technique.  The eventual goal of this work is to produce
  a CCD model validated by laboratory data which can then be used to
  evaluate its effects on weak lensing science.}

\keywords{Photon detectors for UV, visible and IR photons
  (solid-state) (PIN diodes, APDs, Si-PMTs, CCDs, EBCCDs etc);
  Detector modeling and simulations II (electric fields, charge
  transport, multiplication and induction, pulse formation, electron
  emission, etc)}

\begin{document}

\section{Introduction}

A requirement in the design of detectors and associated measurement
algorithms in next generation survey instruments is that the
systematic errors caused by instrumental effects in the image plane
CCDs are not themselves larger than the size of the signals being
measured.  This is especially important in weak lensing analyses which
rely on careful measurements of PSF-convolved galaxy shapes.  A
particular concern in these analyses is the effect of electric fields
in the bulk of the CCDs, which can cause lateral deflections of the
electrons as they drift through the sensor.

In the LSST project~\cite{Ivezic:2008fe} we are using a Monte Carlo approach to
try to simulate physical effects in the sensors.  PhoSim~\cite{phosim,
  phosim-paccd} is a photon-by-photon tracking simulation package for
optical telescopes.  It can follow photons through the atmosphere,
into and through the optics of the telescope and finally into the
image plane itself where the photon's interactions with the silicon
are modeled and the resulting electrons are tracked through the CCD
and collected to form pixelized images.  The goal of this work is to
understand and control the instrument based systematic errors which
might affect the extraction of cosmological parameters in the LSST
science analysis. 

The overall LSST strategy is that, whenever possible, we will always
employ a physics-based model for the effects in the instrument
including electron-by-electron tracking.  Driving this philosophy is a
desire to not trivially simulate and then correct for instrumental
effects with similar parameterized models.  By relying on physics-based
models which we have validated with real data taken in the lab we can
evaluate our correction and measurement  algorithms.

At this point in our studies, we usually turn one instrumental effect
on at a time with everything else turned off. We often magnify the
effects in order for us to understand them and we vary their size in
order to match the observed magnitudes in the laboratory.  Finally,
running and tracking each electron separately can be quite
time-consuming.  For these reasons, time-savings measures are
sometimes employed during the simulation runs.  For example, we
sometimes reduce the size of the simulated CCD chips by up-to a factor
of ten.

In this talk, I will focus on the simulation and validation of some of
the main sensor effects that are currently implemented in PhoSim.
Specifically, I will focus on the so-called Brighter-Fatter effect in
which charge already collected in a pixel is believed to deflect later
arriving drifting electrons into adjacent pixels, thereby modifying
both the expected spot size as a function of height and the expected
relationship between the mean pixel occupancy and it's variance in a
flat illumination. I will also discuss the behavior of charge as it
approaches the edge of the sensor.  Finally, in this talk I will discuss
the work done by Andrei Nomerotski and Ben Beamer on the effect of
``Tree-Ring'' like doping inhomogeneities in the silicon boules.  You
can find this work described in Mr. Beamer's contribution to these
proceedings~\cite{beamer}.

\section{The Brighter-Fatter Effect}
\label{sec:BF}

The Brighter-Fatter (BF) effect is now a well described phenomenon
that has seen by many groups, and several have produced models to
describe it~\cite{downing, antilogus,
  Guyonnet:2015soa,Rasmussen:2014qwa}. Recently, the DECam team has
begun to characterize and and correct for the effect in their
analyses~\cite{Gruen:2015nca,Plazas:2014aha}. As described above, the
BF effect is modeled as a consequence of the space-charge which has
already been collected in the pixels of the CCD.  As the next electron
follows the field lines down to the collection area the charge already
in the pixel deflects the electron horizontally.  One can also think
of this problem as an effective mapping between the effective pixels
on the surface of the CCD and the pixels in the charge collection
area.  The presence of charge in the well can warp the effective pixel
area on the surface, effectively reducing it as the modified field
lines carry the charge to different pixels at the wells.

Importantly, the BF effect can also modify the expected relationship
between the mean value and variance of the electron count in each well
in flat illuminations of the sensor.  Charge that might normally be in
a well is diverted if there is already charge in that well thus
disturbing the expected poisson statistics.  This effect reduces the
observed variance relative to the expectation.
{\it
A well motivated and implemented physical model should correctly
simulate both the spot spreading and flat variance reduction effects
with the same parameters.
}

One of the main insights that came from the preparation of this talk
was that currently, different groups are implementing the modification
of the field lines in the CCD from collected charge differently.
Currently, PhoSim solves for the static electrostatic fields by
solving Poisson's equation for the static charge distributed by the
electronic structure of the CCD including the guard-rails etc.  Then a
dipole field which is supposed to be induced by the collected charge
and a oppositely charged mirror charge in the CCD is superimposed on
this field.  The dipole field changes dynamically as the well fills.
The strength of this dipole (as determined by the dipole length)
governs the strength of the BF effect as implemented in PhoSim. 

Discussion at the meeting revealed that the simple addition of a
dipole field to the electrostatic one was likely non-realistic and
other groups did either a full electrostatic calculation or a more
complete series of image charge calculations in order to determine the
proper field to implement.  Work will soon begin to modify the PhoSim
implementation of this effect.  It should be stressed that the physical
model which describes the BF effect is still being studied by the
community. Although phenomenological models such of that
in~\cite{Guyonnet:2015soa} can fit observations, the true
electrostatics of the system must be better understood and
predictive models must be compared with lab-bench data.

\subsection{Spots}

In order to test the BF effect on a spot like source all other effects
are turned off in PhoSim and a Gaussian spot with a width tuned to
data sets at low light level is produced.  A spot calibrated in
photoelectron level is produced by determining the relationship
between the PhoSim sky magnitude and the number of collected electrons
on the CCD.  Here a gain of one is assumed and the raw
``electron-level'' data is analyzed.

Spot intensities ranging from a few thousand electrons to 100,000
electrons are generated with a flat SED and then their width is
extracted using the Sloan Digital Sky Survey shape measurement
algorithm as implemented in the LSST software
stack~\cite{LSST-stack,Bernstein:2001nz}.  The BF effect should cause
the width of the spot to grow with it's intensity. This procedure is
repeated for for many BF parameters ranging from no BF effect at all
to 500 times the nominal PhoSim BF effect.  The strength of the BF
effect is modified by changing the length of the of the dipole induced
by the extra charge in the model.

\begin{figure}[tbp]
  \centering
  \includegraphics[width=.8\textwidth]{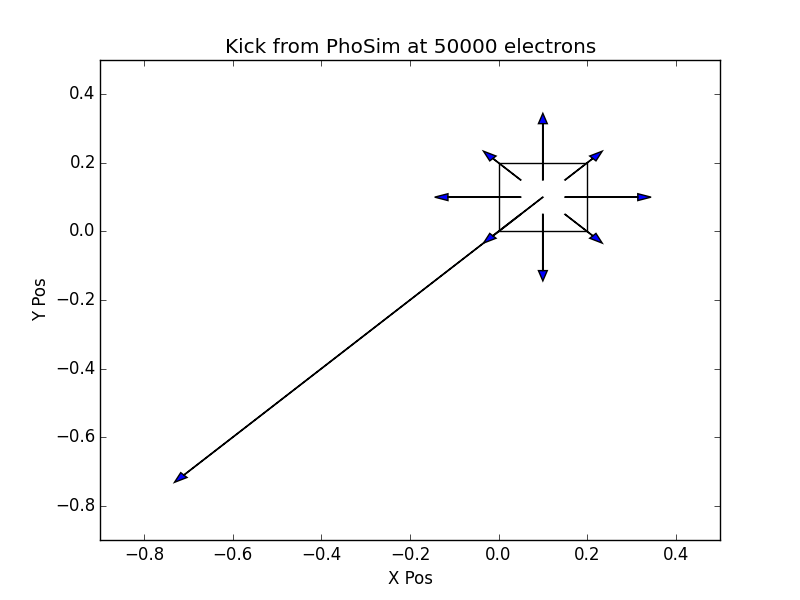}
  \caption{The kick in arcseconds due to the charge held in the center
    of a pixel imparted to an incoming electron when 50,000 electrons
    have already been collected.  Each pixel (represented by the box
    in this figure) is 0.2'' on a side in the LSST image plane.  In
    PhoSim, the existing charge is modeled as being exactly in the
    center of the pixel.  The nine arrows represent the kick imparted
    from the field of that charge to an electron at nine different
    points on the pixel.  The field from other pixels and static
    charge in the CCD structure must also be included to determine the
    net displacement.}
  \label{fig:movement}
\end{figure}

Currently, the PhoSim model assumes that all charge is held in the
exact center of the pixel.  Figure~\ref{fig:movement} shows the effect
of this assumption on the lateral kick given to the drifting electron
with arbitrary normalization from only the  charge at the center of
the pixel at nine different positions in the pixel.  As can be seen
charge drifting exactly to the center of the pixel (where the current
charge is located) feels the largest effect.  

\begin{figure}[tbp]
  \centering
  \includegraphics[width=.9\textwidth]{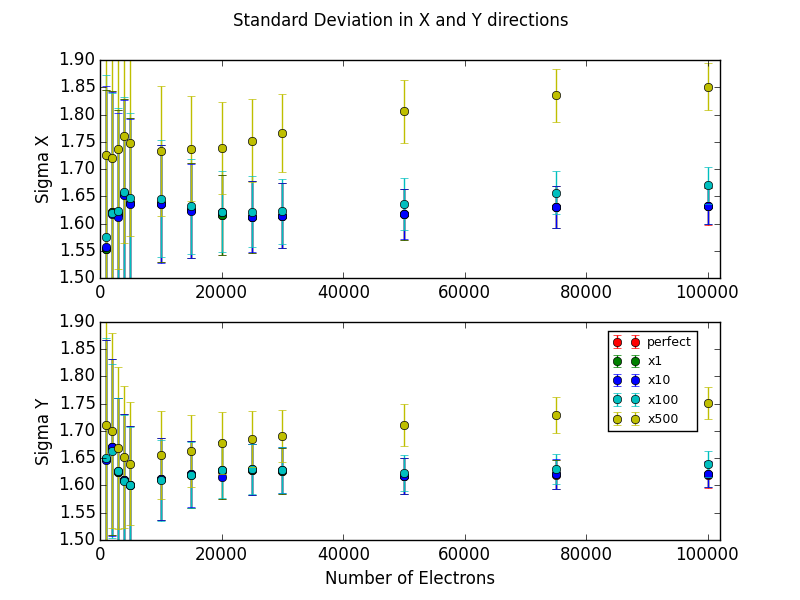}  
  \caption{The measured X and Y width of the gaussian spot in pixels
    as a function of electron level for four different BF parameters
    ranging from no effect to 500 times the nominal PhoSim effect.
    The incoming Gaussian spot has a width of 1.6 pixels.}
  \label{fig:spots}
\end{figure}

In~\cite{antilogus} Astier et al, found in data that the width of the
spot grew by 2-3\% over a similar range of intensities.  The result of
the phoSim simulation are shown in figure~\ref{fig:spots}.  In this
figure, the widths in the X and Y direction are shown separately and
the selected BF parameters range from no effect at all (``Perfect'')
to 500 times the nominal.  As can be seen, in order to obtain spot
spreading at the 5\% level, the size of the effect in the model must be
increased to the order of 100 times the nominal.

The differences between the data and simulation in the case may be due
to the unrealistic modeling of a pure dipole located exactly at the
center of the pixel.  We plan to modify the model of the charge and
see if this improves the agreement.

\subsection{Flats}

As described in section~\ref{sec:BF}, the BF effect should also affect
the relationship between the variance and the mean of each
pixel. Generally the ratio of these two quantities as a function of
intensity is known as the Photon Transfer Curve (PTC) and can be used
in the linear regime to extract the gain of the system.
Non-linearities induced in the PTC by the BF effect was first
described in detail in~\cite{downing}. The authors of~\cite{downing}
also pointed out that if the non-linearity of the PTC was due to
charge moving from one pixel to another then there should also be
non-zero correlation coefficients between pixels, and further that
grouping together pixels into larger blocks should mitigate the
effect.  In~\cite{antilogus} this effect was measured in e2v LSST CCDs
and a correlation coefficient of 1 to 5\% was found to a pixel's
nearest neighbors.  A larger coefficient was found in the y-direction
presumably due to the presence of charge-stops impeding the transfer
of charge in the x direction.

In order to test for this effect in PhoSim, I once again simulated
calibrated levels of light on a photon-by-photon basis.  In this case
a flat illumination at 550 nm was simulated with electron levels
ranging between 100 and 200,000 electrons per pixel for a set of BF
strengths.  All photon level optimizations are turned off in the
simulation and so, in order to reduce computational run-time, the size
of the simulated CCD is reduced from 4000x4000 to 400x400 pixels.
Also, in order to remove any mean variations across the CCD plane
caused by other effects (such as edge roll-off) two exposures were
simulated and subtracted as in real data exposures, thus removing any
common-mode effects.

\begin{figure}[tbp]
  \centering
  \begin{minipage}{4.0in}
    \includegraphics[width=4.0in]{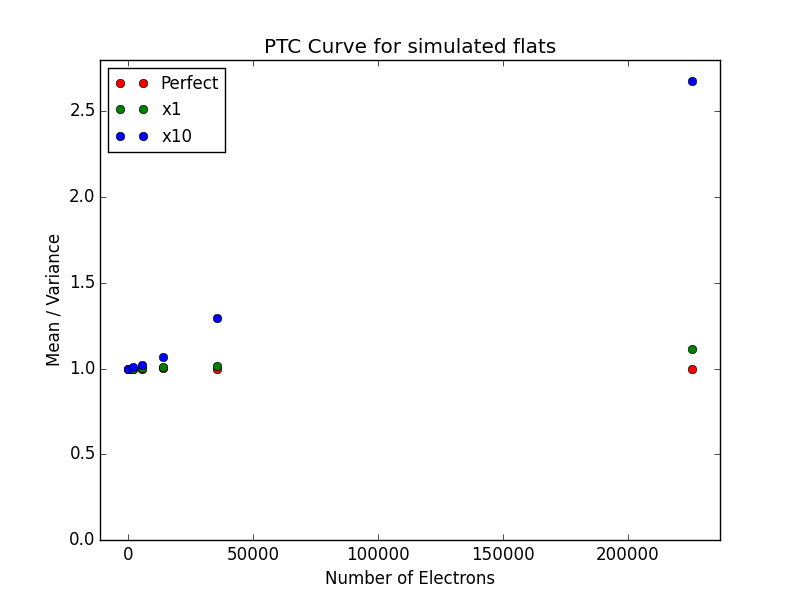}
  \end{minipage}
  \begin{minipage}{4.0in}
    \includegraphics[width=4.0in]{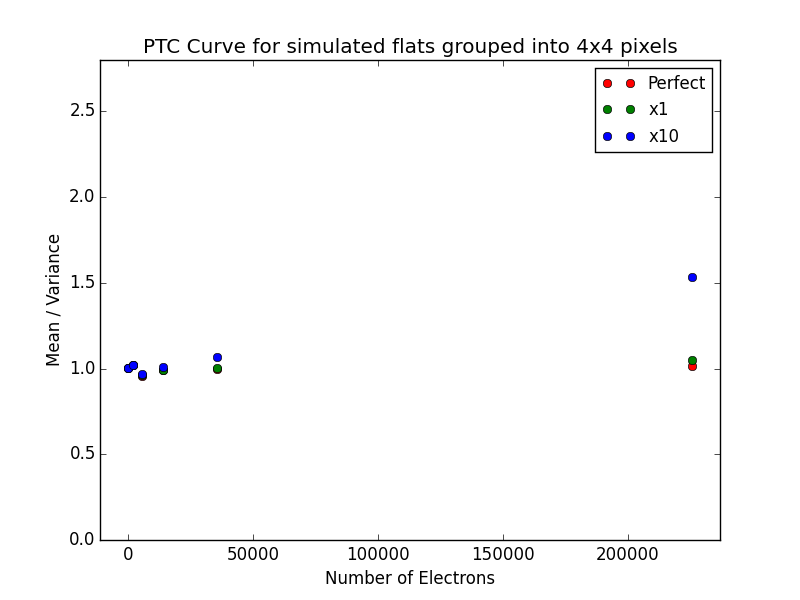}
  \end{minipage}
  \caption{(Top) the common mode subtracted mean/variance as a function of
    electron level for three different BF parameters.  (Bottom) the
    same result with pixels grouped into larger 4x4 blocks. }
  \label{fig:PTC}
\end{figure}

First the PTC curve was examined.  Figure~\ref{fig:PTC} shows the
common-mode subtracted mean/variance for with no BF effect applied
along with the curve obtained with the nominal BF strength and ten
times the normal dipole strength.  As can be seen, with no effect the
variance scales perfectly with the mean.  The expected effect can be
easily seen at one and ten times the nominal effect and, in the bottom
panel, the results for the same analysis with the pixels ganged into
4x4 super-pixels is shown.  As expected the size of the effect is
reduced.

In order to qualitatively compare the results of the simulation with
that of data reported in~\cite{antilogus} I next calculated the spatial
auto-correlation coefficient between each pixel and it's nearest
neighbors on the subtraction of the two flat exposures.  I then report
horizontal correlation coefficient which is the average of the two
nearest horizontal neighbor coefficients and a vertical coefficient
which it the average of the two nearest neighbor vertical
coefficients.   It should be noted that the correlation coefficient
in this case is not position dependent but rather is the statistical
calculation of the correlation coefficient of every pixel with it's
neighbors on one CCD.

\begin{figure}[tbp]
  \centering
  \begin{minipage}{4.0in}
    \includegraphics[width=4.0in]{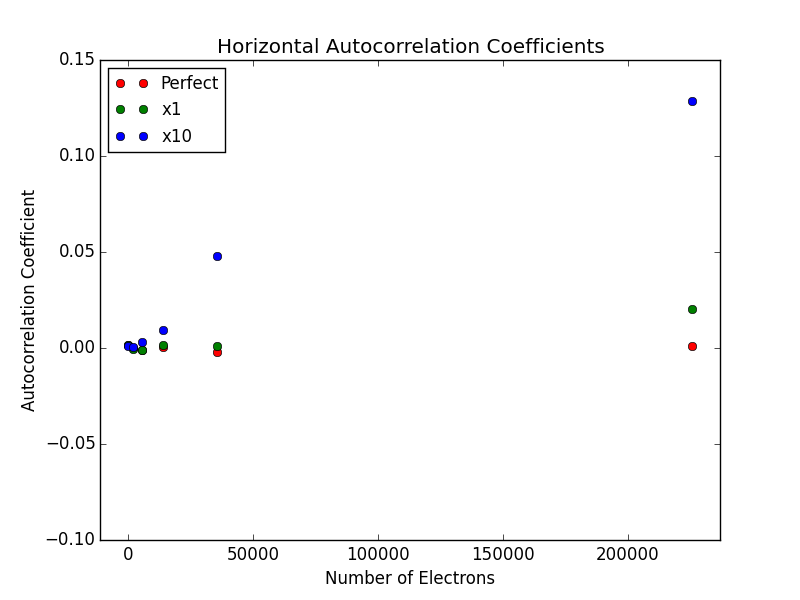}
  \end{minipage}
  \begin{minipage}{4.0in}
    \includegraphics[width=4.0in]{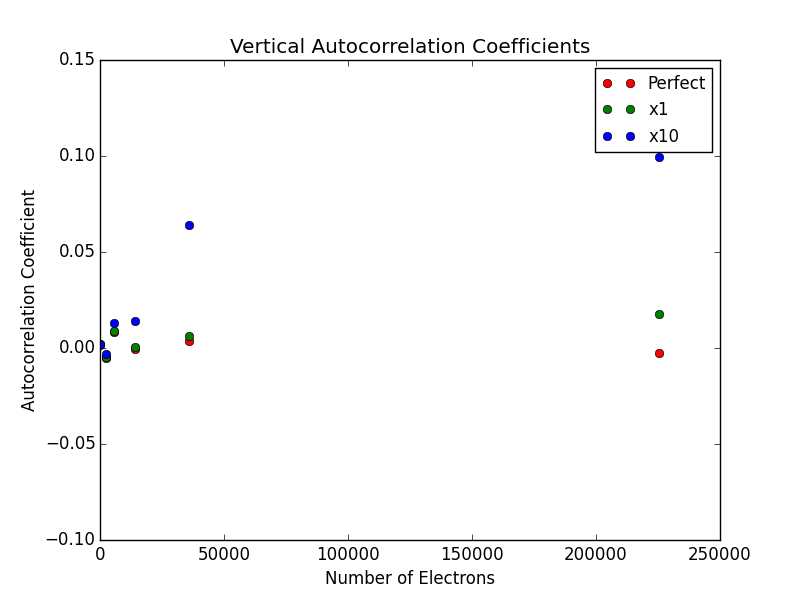}
  \end{minipage}
  \caption{The horizontal (top) and vertical (bottom) auto-correlation
    coefficients between neighboring pixels for common mode
    subtracted pairs of exposures as a function of electron level for
    three different BF parameters.}
  \label{fig:correlation}
\end{figure}

The results of the simulation are shown in
Figure~\ref{fig:correlation} for both the horizontal and vertical
coefficients.  As expected with no BF effect, no correlation is seen.
However at the nominal BF dipole magnitude we see about a 2\% effect.
This is very close to the size of the effect measured in data
by~\cite{antilogus}.  

\subsection{Brighter Fatter Effect Conclusions}

Somewhat surprisingly, we find that while the BF effect does not seem to be
large enough when considering spot data compared to the previously
reported data result in~\cite{antilogus}, it seems to match the data
results well when measuring auto-correlation coefficients.  

The first and most important caveat to consider is that the same
algorithms I used to analyze the simulation data should be used to
analyze real data.  This is the only way to make a fair comparison.
However, if the discrepancy between data and simulation still persists
in that case it may be that the incomplete dipole model currently
employed in PhoSim is the cause.  More studies are necessary.

\section{ Edge Effects}

\begin{figure}[btp]
  \centering
  \includegraphics[width=.7\textwidth]{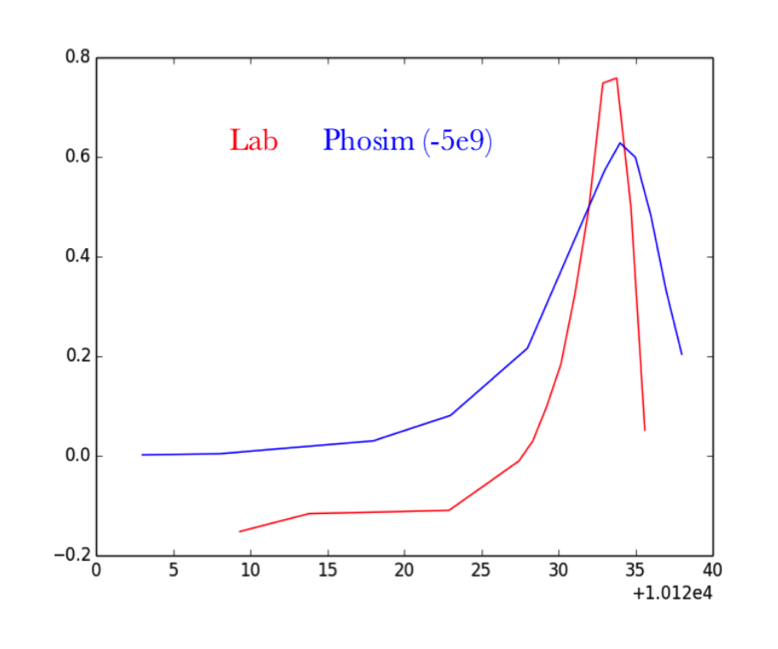}    
  \caption{The astrometric shift as a function of linear position as a spot
    moves towards the edge of the sensor for both lab data (red) and
    PhoSim output (blue).}
  \label{fig:edge}
\end{figure}

Due to the presence of the guard rails in the CCD, electrons near the
edge of the sensor feel a force which pulls them towards the edge.
This manifests itself in a roll-off effect and an astrometric shift.
We tested this effect in Phosim by making a grid of simple stars and
moving them towards the edge of the sensor.  Then, the sextractor
package~\cite{Bertin:1996fj} was used to measure their positions, and the
astrometric shift was measured by comparing their true and measured
positions.  Figure~\ref{fig:edge} shows a comparison of the simulation
data with lab data taken at BNL.
These shapes are not the same which suggests that the surface charge
contributions need to be tuned. The DECam's edge effect also does not entirely follow
the behavior as expected by their model~\cite{Plazas:2014aha} which
also implies that a better description of the charge distributions
near the edge of the sensors is needed.  

Additionally, in the process of doing the BF study I discovered a
roll-off effect due to the BF effect itself.  Namely, at least in the
way we simulate the effect, it is possible for charge to be pushed off
the edge of the chip by the fields present from already collected
charge.  On the other hand, this process does not work in the other
direction.  Due to the guard rail, charge from outside the sensitive
region does not get pushed in.  So, as you move towards the edge on a
flat exposure, some of the charge is pushed out of the sensitive
region resulting in a roll off.
 
\begin{figure}[tbp]
  \centering
  \includegraphics[width=.75\textwidth]{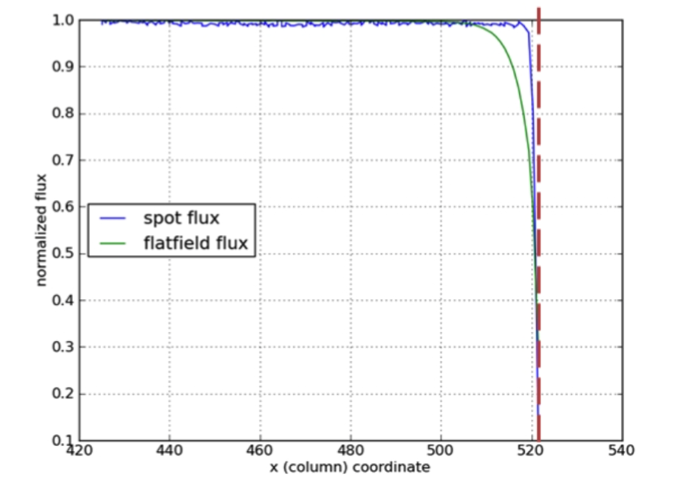}    
  \caption{The decrease in measured flux as a function of CCD column
    coordinates for both a spot (blue) and a flat field (green). }
  \label{fig:rolloff}
\end{figure}

Figure~\ref{fig:rolloff} show a comparison between the roll-off of
spot data and flats in lab data taken at BNL.  They behave
differently.  It is possible this is related to the BF related
roll-off described above. We need to further simulate and test these
two configurations in order to draw conclusions.

\section {Tree Rings}

In the multi-day process of growing the silicon boules used to produce
CCD sensors the doping intensity may vary as a function of time and
thus position across the boule.  The variations resemble
``tree-rings'' in their morphology and this gradient in doping
concentration induces lateral fields inside the CCD which can distort
shapes of imaged objects as the electrons drift down to the charge
buckets.  Descriptions of the simulation of this effect in PhoSim can
be found elsewhere in these proceedings~\cite{beamer}.

\section {Conclusions}

We have begun a concerted research program to simulate and validate
physics CCD effects using  PhoSim and real data.  Currently, the
tree-ring simulation seems to be working well and we will now tune the
parameters.  A first-pass model for the science groups should be
avaliable within a few months of the publication of these
proceedings.  

On the other-hand, it seems as if the model we are using to simulate
the BF effect still needs further work.  Direct comparisons with lab
bench data will happen soon and should help us to refine the model.
Other effects in the CCDs will be explored after this work is
completed.  Eventually we will use this model to estimate the impact
on weak lensing shear analysis from these instrumental systematics.

\acknowledgments

The author would like to acknowledge and thank Andrei Nomerotski,
Benjamin Beamer and Max Duncan for contributions on the tree-ring and
edge effects studies, John Peterson for his work and help in using the
PhoSim package, and Robert Lupton for useful conversations regarding
the overall LSST simulation and measurement strategy.

The author would also like to thank the PhoSim and LSST teams for
general help with the LSST software stack, and Duke University and the
DOE Office of Science for supporting this research.


\begin{thebibliography}{9}


\bibitem{Ivezic:2008fe} 
  Z.~Ivezic  et al.  [LSST Collaboration],
  \emph{LSST: from Science Drivers to Reference Design and Anticipated Data Products},
  arXiv:0805.2366 [astro-ph].

\bibitem{phosim}
J.R. Peterson et al., 
\emph {Simulation of Astronomical Images from Optical Survey
  Telescopes Using a Comprehensive Photon Monte Carlo Approach},
In preparation (2015)

\bibitem{phosim-paccd}
J.R. Peterson, 
\emph {PhoSim: a code to simulate telescopes one photon at a time},
\emph{JINST} \textbf{9} C04010

\bibitem{beamer}
B. Beamer, 
\emph {A study of astrometric distortions due to ``tree rings' in CCD
  sensors using the LSST Photon Simulator},
These Proceedings

\bibitem{downing}
Mark Downing, Dietrich Baade, Peter Sinclaire, Sebastian Deiries,
Fabrice Christen,
\emph{CCD riddle: a) signal vs time: linear; b) signal vs variance:
  non-linear},
\emph{Proc. SPIE 6276, High Energy, Optical, and Infrared Detectors
  for Astronomy II}, 627609 (June 15, 2006)

\bibitem{antilogus}
P Antilogus {\it et al},
\emph{The brighter-fatter effect and pixel correlations in CCD
  sensors}
\emph{JINST} \textbf{9} C03048 

\bibitem{Guyonnet:2015soa} 
  A.~Guyonnet, P.~Astier, P.~Antilogus, N.~Regnault and P.~Doherty,
  \emph{Evidence for self-interaction of charge distribution in charge-coupled devices},
  Astron.\ Astrophys.\  {\bf 575}, A41 (2015)
  [arXiv:1501.01577 [astro-ph.IM]].

\bibitem{Rasmussen:2014qwa} 
  A.~Rasmussen, P.~Antilogus, P.~Astier, C.~Claver, P.~Doherty, G.~Dubois-Felsmann, K.~Gilmore and S.~Kahn {\it et al.},
  \emph{A framework for modeling the detailed optical response of thick, multiple segment, large format sensors for precision astronomy applications},
  Proc.\ SPIE Int.\ Soc.\ Opt.\ Eng.\  {\bf 9150}, 915017 (2014)
  [arXiv:1407.5655 [astro-ph.IM]].


\bibitem{Gruen:2015nca} 
  D.~Gruen, G.~M.~Bernstein, M.~Jarvis, B.~Rowe, V.~Vikram, A.~A.~Plazas and S.~Seitz,
  \emph{Characterization and correction of charge-induced pixel shifts in DECam},
  These Proceedings, arXiv:1501.02802 [astro-ph.IM].

\bibitem{Plazas:2014aha} 
  A.~A.~Plazas, G.~M.~Bernstein and E.~S.~Sheldon,
  \emph{Transverse electric fields' effects in the Dark Energy Camera CCDs},
  JINST {\bf 9}, C04001 (2014)
  [arXiv:1403.6127 [astro-ph.IM]].

\bibitem{LSST-stack}
T.~Axelrod, J.~Kantor, R. H.~Lupton and F.~Pierfederici,
\emph{An open source application framework for astronomical imaging
  pipelines}, 
Proc. SPIE 7740, Software and Cyberinfrastructure for Astronomy,
774015 (2010)

\bibitem{Bernstein:2001nz} 
  G.~M.~Bernstein and M.~Jarvis,
  \emph{Shapes and shears, stars and smears: optimal measurements for weak lensing},
  Astron.\ J.\  {\bf 123}, 583 (2002)
  [astro-ph/0107431].

\bibitem{Bertin:1996fj} 
 E.~Bertin and S.~Arnouts,
\emph{SExtractor: Software for source extraction},
 Astron.\ Astrophys.\ Suppl.\ Ser.\  {\bf 117}, 393 (1996).

\end{thebibliography}
\end{document}